\documentclass[seceq]{ptptex}
\usepackage{wrapft}

\newcommand{\ave}[1]{\langle #1 \rangle}

\usepackage{amssymb}
\usepackage[dvips]{graphicx}
\usepackage{booktabs}
\usepackage{tabularx}
\usepackage{graphicx,color} 
\usepackage{wrapft}

\notypesetlogo
{
\protect

\title{Particle and Energy Transport in quantum disordered and quasi-periodic chains connected to mesoscopic Fermi reservoirs}
}

\author{
Shigeru \textsc{Ajisaka}\footnote{Email: g00k0056@suou.waseda.jp},
Felipe \textsc{Barra},
Carlos \textsc{Mej\'ia-Monasterio}
Toma\v{z} \textsc{Prosen}
}

\inst{
Departamento de F\'isica, Facultad de Ciencias F\'isicas y
Matem\'aticas, Universidad de Chile, Casilla 487-3, Santiago Chile
\\
Laboratory of Physical Properties, Technical University of Madrid, Av. Complutense s/n 28040, Madrid, Spain
\\
Faculty of Mathematics and Physics, University of Ljubljana, Jadranska 19, SI-1000 Ljubljana, Slovenia
}

\abst{

  We study a model of nonequilibrium quantum transport of particles
  and energy in a many-body system connected to mesoscopic Fermi
  reservoirs (the so-called meso-reservoirs).  We discuss the
  conservation laws of particles and energy within our setup as well
  as the transport properties of quasi-periodic and disordered chains.

}

\begin{document}
\maketitle

\section{Introduction}

Understanding the macroscopic transport from a microscopic point of view
is a central topic of statistical physics. Especially, the development of
nanoscale devices reveals unconventional transport.  Thus, the study
of transport in mesoscopic systems has significant value both in
fundamental theory and for applications in future technologies.  Since
mesoscopic systems are strongly coupled with the environment, their
understanding in nonequilibrium regimes requires knowledge of the global
features of the total system including the environment (reservoirs).
Recently, we proposed a model that comprises mesoscopic reservoirs
with a finite number of degrees of freedom~\cite{ABMP12} (the
so-called meso-reservoirs), and study the transport properties of
periodic chains. In particular, the parameter dependence of the
transport properties as well as the Onsager reciprocity relation were
studied.

In this paper, we study periodic chains in more detail, as well as
quasi-periodic and disordered chains.  We will also discuss the
conservation law of energy and particles within our setup.

\section{Model}

A key idea of our model is to enforce the finite reservoirs to
equilibrium (or almost equlibirum) state using the Lindblad
dissipator\cite{Koss76,Lindblad76}.  In particular, if the term in the
Liouvillean evolution containing the Lindblad dissipator is small, we
can interpret that the `traced-out-infinite-reservoirs' ({\it
  super-reservoirs}) force our finite reservoirs ({\it
  meso-reservoirs}) to equilibrium.  That is to say, our density
matrix follows the Lindblad equation of the following form:
\begin{eqnarray}
\frac{{\rm d} \rho}{{\rm d} t} &=&
-i[H,\rho] + D(\rho) \\
D(\rho) &=&
\sum_{k, \alpha, m}
\left(
2L_{k, \alpha, m} \rho L_{k, \alpha, m}^\dag
-\{L_{k, \alpha, m}^\dag L_{k, \alpha, m},\rho\}
\right) \nonumber
\\
H &=& H_S+H_L+H_R+V
\nonumber
\nonumber
\\
H_S &=& -\sum_{j=1}^{n-1} \Big(t_j c_j^\dag c_{j+1} + (h.c.) \Big)
+\sum_{j=1}^{n} U_j c_j^\dag c_j
\nonumber
\\
H_{\alpha} &=& \sum_{k=1}^{K} 
\epsilon^\alpha_k a_{k\alpha}^\dag a_{k\alpha}
,\ \ \alpha=L, R
\nonumber
\\
V &=& \sum_{k=1}^{K} \left( 
v^L_k a_{kL}^\dag c_1 + v^R_k a_{kR}^\dag c_n\right) + (h.c.)
\nonumber
\\
L_{k,\alpha,1} &=& \sqrt{ \Gamma_{k,\alpha,1} } a_{k\alpha}
,\ \
L_{k,\alpha,2}=\sqrt{ \Gamma_{k,\alpha,2} } a_{k\alpha}^\dag
\nonumber
\\
\Gamma_{k,\alpha,1}&=&\gamma^\alpha_k ( 1 - F_{\alpha}(\epsilon_k) )
,\ \ 
\Gamma_{k,\alpha,2}=\gamma^\alpha_k F_{\alpha} (\epsilon_k)
\ \ ,
\nonumber
\end{eqnarray}
where $c_j$ is the annihilation operator of system fermions,
$a_{k,\alpha}$ is that of reservoir fermions with wave number $k$,
$t_j$ is the nearest neighbor hopping, $U_j$ is the onsite potential,
$v^\alpha_k$ is the coupling between the system and reservoirs, and
$F_\alpha(\epsilon)=(e^{\beta_\alpha (\epsilon-\mu_\alpha)}+1)^{-1}$
are Fermi distributions, with inverse temperatures $\beta_\alpha$ and
chemical potentials $\mu_\alpha$, while $[\cdot , \cdot]$ and $\{
\cdot , \cdot\}$ denote the commutator and anti-commutator,
respectively.  The parameter $\gamma^\alpha_k$ determines the strength
of the coupling to the super-reservoirs and needs to be fine-tuned in
order to ensure the applicability of the model~\cite{ABMP12}.  We
stress that our model does not rely on the usual weak-coupling
assumption needed for the physical derivation of the Lindblad master
equation \cite{Breuer02}, thus $\gamma^\alpha_k$ do not need to be
small parameters.

\section{Conservation laws}
\subsection{Particle conservation}
In this section, we discuss conservation of particle number and
energy.  We focus first on the particle current.  Let $n_j=c^\dag_j
c_j$ and $n^\alpha_k=a_{k\alpha}^\dag a_{k\alpha}\ (\alpha=L,R)$,
then, with time derivatives that can be casted as 
\begin{eqnarray}
\frac{d n_j }{dt} &=&  J^P_{j-1} -J^P_{j},\ (1\le j\le n-1)
\nonumber
\\
J^P_0 &\equiv& i \sum_k v^L_k (a^\dag_{kL} c_1 - c_1^\dag a_{kL})
\equiv \sum_k j^L_k
\nonumber
\\
J^P_{j}&\equiv & -i t_{j}(c^\dag_{j} c_{j+1} - c^\dag_{j+1} c_{j})
,\ (1\le j\le n-1)
\nonumber
\\ 
J^P_n &\equiv& -i \sum_k v_k^R (a^\dag_{kR} c_n - c_n^\dag a_{kR})
\equiv \sum_k j^R_k
\nonumber
\\ 
\frac{d n^L_k }{dt} &=& -j^L_k + D(n^L_k)
\nonumber
\\ 
\frac{d n^R_k }{dt} &=& j^R_k + D(n^R_k)
\ \ ,
\end{eqnarray}
where the dissipative parts~$D(n^\alpha_k)$ are given by
\begin{eqnarray}
D(n^\alpha_k) &=& 
-2\Gamma_{k,\alpha,1}\ 
n^{\alpha}_k
+
2\Gamma_{k,\alpha,2}
( {\bf 1}-n^\alpha_k)
\nonumber
\\
&=& 
-2\gamma^\alpha_k n^{\alpha}_k
+
2\gamma^\alpha_k F_\alpha(k) {\bf 1}
\ \ .
\end{eqnarray}
By taking the expectation value in the {\it nonequilibrium steady
  state} (NESS) $\ave{\cdot}={\rm tr} (\cdot \rho(t\to\infty))$, we
obtain
\begin{eqnarray}
\sum_k \ave{j^L_k} &=&\ave{J_0}=\ave{J_1}=\cdots=\ave{J_{n-1}}
=\ave{J_n}=\sum_k \ave{j^R_k}\equiv J^P
\nonumber
\\
\ave{j_k^L}&=&\ave{D(n^L_k)},\ \ 
\ave{j_k^R}=-\ave{D(n^R_k)}\ \ .
\end{eqnarray}
Thus, the particle current from the left meso-reservoir to the
system~$\sum_k \ave{j^L_k}$ is equal to the current from the system to
the right meso-reservoir~$\sum_k \ave{j^R_k}$, and there is no
particle loss due to the existence of the Lindblad dissipators.
Moreover, we have
\begin{eqnarray}
J^P=\sum_k\ave{D(n^L_k)} = -\sum_k\ave{D(n^R_k)} 
,\ 
\end{eqnarray}
and it follows
\begin{eqnarray}
J^P &=& 2\sum_k \gamma^L_k \left\{  \ave{n^{L}_k} - F_L (k) \right\} 
=
-2\sum_k \gamma^R_k \left\{ \ave{n^{R}_k} - F_R(k)  \right\} \ .
\label{particle_Landauer}
\end{eqnarray}
This expression can be understood as a generalization of the
Landauer formula for the current between the meso-reservoirs and the
super-reservoirs.  Moreover, if $\gamma^\alpha_k$ is independent of
$k$, the total differences of charge densities from their equilibrium
state values are proportional to $(1/\gamma^\alpha)$.

\subsection{Energy conservation}
We now turn our attention to the  conservation of energy.
Casting the local energy density as
\begin{eqnarray}
H_j &\equiv& -\Big(t_j c_j^\dag c_{j+1} + t_j c_{j+1}^\dag c_j \Big)
+U_j c_j^\dag c_j
\ , (1\le j\le n)
\nonumber
\\
c_{n+1} &\equiv& 0
\ ,
\end{eqnarray}
their time derivatives are
\begin{eqnarray}
\frac{d H_j }{dt} &=& 
J_{j-1}^E-J^E_{j},\ (1\le j\le n-2)
\nonumber
\\
J^E_{0} &\equiv& U_1 J^P_0+i\sum v^L_kt_1(c_2^\dag a_k-a_k^\dag c_2)
\nonumber
\\
J^E_{j} &\equiv& it_j t_{j+1}(c^\dag_j c_{j+2} - c^\dag_{j+2} c_j)+U_{j+1} J_j^P
,\ (1\le j\le n-2)
\nonumber
\\ 
\frac{d H_{n-1} }{dt} &=& J_{n-2}^E-J^E_{n-1}-\widetilde{J}^E_n
\nonumber
\\  
J_{n-1}^E &\equiv&   U_n J^P_{n-1}
\nonumber
\\  
\widetilde{J}_{n}^E &\equiv&   i\sum_k v^R_k t_{n-1}
(b^\dag_k c_{n-1} - c^\dag_{n-1} b_k)
\nonumber
\\ 
\frac{d H_{n} }{dt} &=& J^E_{n-1}-J^E_n
\nonumber
\\  
J_{n}^E &\equiv&   UJ^P_{n}
=
iU \sum_k v_k^R (c_n^\dag a_{kR} - a^\dag_{kR} c_n)
\nonumber
\\
\frac{d V_L }{dt} &=& -J^E_0+J^E_{L\to V}+D(V_L)
\nonumber
\\
\frac{d V_R }{dt} &=& J^E_n+\widetilde{J}^E_n- J^E_{V\to R}+D(V_R)
\nonumber
\\  
J^E_{L\to V} &\equiv&   i\sum \epsilon_k^L v_k^L (a^\dag_k c_1-c_1 a_k\dag)
\equiv \sum_k \epsilon_k^L j^L_{k}
\nonumber
\\  
J^E_{V\to R} &\equiv&   -i\sum \epsilon_k^R v_k^L (b^\dag_k c_n-c_n b_k\dag)
\equiv \sum_k \epsilon_k^R j^R_{k}
\nonumber
\\
\frac{d H_L }{dt} &=& -J^E_{L\to V}+D(H_L)
\nonumber
\\
\frac{d H_R }{dt} &=& J^E_{V\to R}+D(H_R)\ \ ,
\end{eqnarray}
where the dissipative terms $D(V_\alpha)$ and $D(H_\alpha)$ are given
by
\begin{eqnarray}
D(V_L) &=& -\sum_k \gamma^L_k v^L_k (a_{kL}^\dag c_1+c_1^\dag a_{kL})
\nonumber
\\
D(V_R) &=& -\sum_k \gamma^R_k v^R_k (a_{kR}^\dag c_n+c_n^\dag a_{kR})
\nonumber
\\
D(H_\alpha) &=& 2\sum_k \gamma^\alpha_k \epsilon_k (F_L(k){\bf 1}-n^\alpha_k)
\ \ .
\end{eqnarray}
By taking average with respect to the NESS we obtain
\begin{eqnarray}
&& \ave{J_0^E} = \ave{J^E_1}=\cdots=\ave{J^E_{n-2}}
=\ave{J^E_{n-1}} + \ave{\widetilde{J}^E_{n}}
=\ave{J^E_{n}} + \ave{\widetilde{J}^E_{n}}
\nonumber
\\
&& i\sum_k v^L_k t_1 \ave{ c_2^\dag a_k - a_k^\dag c_2 }
= i\sum_k t_{n-1} v^R_k \ave{ a_{kR}^\dag c_{n-1}-c_{n-1}^\dag a_{kR} }
\nonumber
\\
&&
\ave{D(H_L)}+\ave{D(V_L)} =-\ave{D(H_R)}-\ave{D(V_R)}
\ \ .
\end{eqnarray}
Although, the total particle current at the two meso-reservoirs is conserved, 
i.e., $\sum_k \ave{j^L_k} = \sum_k \ave{j^R_k}$, 
the particle current distributions are not the same, i.e.,  
\begin{eqnarray}
 -2\gamma_k^L \{ \ave{n^L_k} - F_L(k)\}=\ave{j^L_k} 
\neq \ave{j^R_k}=
 2\gamma_k^R \{\ave{n^R_k} - F_R(k)\}
\ .
\end{eqnarray}
Thus, the ingoing energy current at the left hand sice is not equal 
to the outgoing energy current at the right hand side, i.e., 
$\sum_k \epsilon_k^L \ave{j^L_k} \neq \sum_k \epsilon_k^R \ave{j^R_k}$.

In our case, 
\begin{eqnarray}
\ave{D(H_L)} &=& \sum_k \epsilon_k \ave{j_k^L}=
2\sum_k \epsilon_k v^L_k\ 
\ {\rm Im} \ave{c_1^\dag a_k}
\end{eqnarray}
and
\begin{eqnarray}
\ave{D(V_L)} &=& - 2\sum_k \gamma^L_k v^L_k\ 
\ {\rm Re} \ave{c_1^\dag a_k}
\end{eqnarray}
are not conseved separately, and it induces the difference of the
energy current from left meso-reservoir to system and that from system
to right meso-reservoir.  Namely, the following four statements are
deeply connected:
\begin{itemize}
\item[(1)] Different amount of energy is dissipated at left and
  right~$\ave{D(V_L)}\neq -\ave{D(V_R)}$.

\item[(2)] The energy current from left meso-reservoir and that to
  right meso-reservoir are different~$\ave{J^E_{L\to S}}\neq
  \ave{J^E_{S\to R}}$.

\item[(3)] The particle current distribution of two reservoirs are
  different~$\ave{j_k^L} \neq \ave{j_k^R}$.

\item[(4)] The particle distribution of two reservoirs are
  different~$\ave{n_k^L}\neq\ave{n_k^R}$.
\end{itemize}

\section{Numerical results}

As discussed in~\cite{ABMP12}, expectation values with respect to the
NESS are easily obtained by solving $2n\times 2n$ dimensional
Sylvester equation.  In this section, we discuss the transport
properties of periodic, Fibonacci and disordered chains at NESS.

We study monoatomic ($t_j=t$), diatomic ($t_{2j-1}=t_A,\ t_{2j}=t_B$),
Fibonacci, and disordered chains.  The Fibonacci chain is constructed
by first taking $n=3$ and setting $t_1=t_A,\ t_2=t_B$ constituting the
first generation, and then inductively replacing $t_A$ by $t_A t_B$
and $t_B$ by $t_A$.  For instance, the second generation yields $n=4$
and $t_1=t_A,\ t_2=t_B,\ t_3=t_A$; the third generation yields $n=6$
and $t_1=t_A,\ t_2=t_B,\ t_3=t_A,\ t_4=t_A,\ t_5=t_B$, and so on.
Alternatively, one can construct $n$-th generation by concatenating
$(n-2)$-th generation after $(n-1)$-th generation.

For disordered chains, we take $t_j$ from a uniform distribution in
$[t-\delta, t+\delta]$.

To make things simple, we have set $U_j=U$, $v^L_k=v^R_k=v$ and
$\gamma^L_k=\gamma^R_k=\gamma$ for numerical results.  As discussed in
\cite{ABMP12}, $v$ should be smaller than $\gamma$, and we have set
$\gamma=0.1,\ v=0.03,\ \epsilon_1=-20,\ \epsilon_K=20,\ K=200$, and
$\mu_L=-\mu_R=\mu$, unless specified differently.

\begin{figure}
\includegraphics[width=135mm]{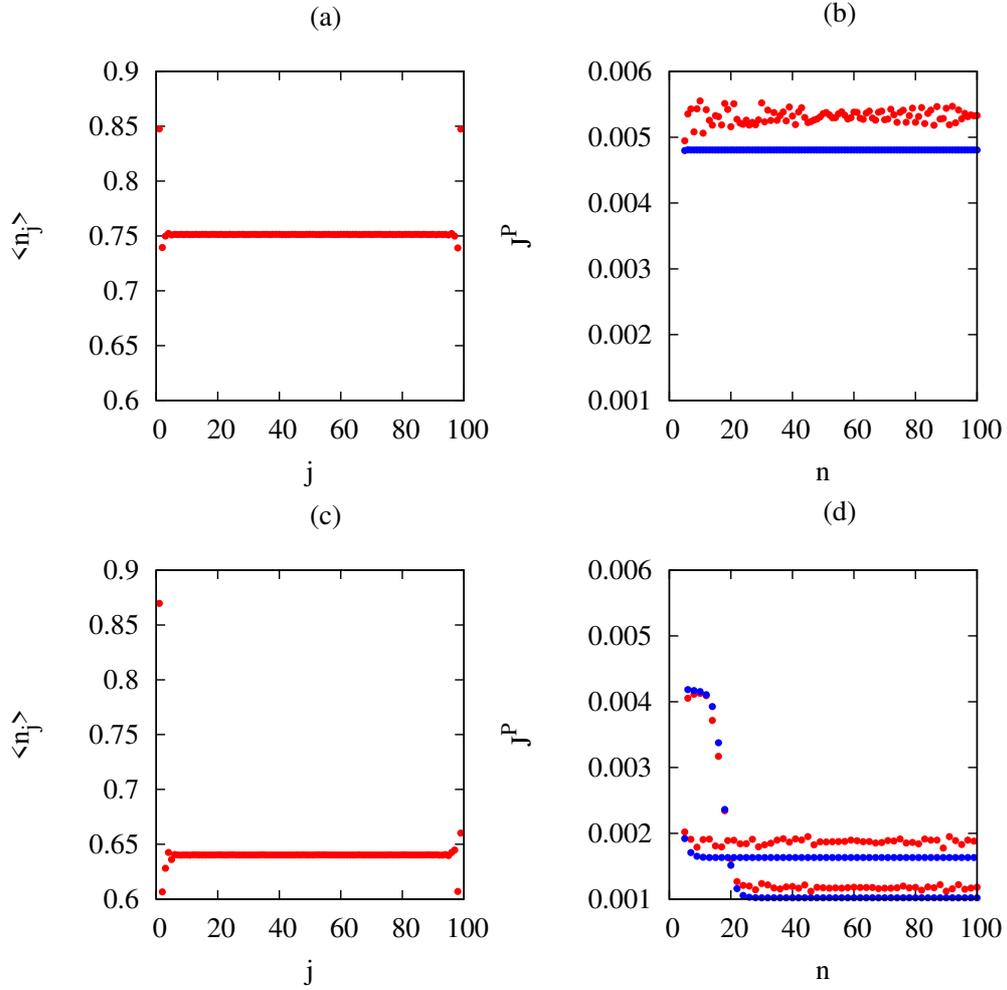}
\caption{\label{periodic} (Color online) Panel~(a) and (c) show
  position ($j$) dependence of occupation density~$n_j$ for (a)
  monoatomic chain, and (c) diatomic chain.  Panels~(b) and (d) show
  $n$ dependence of the particle current~$J^P$ for (b) monoatomic
  chain, and (d) diatomic chain (red: $\gamma=0.1$, blue: $\gamma=1$)
}
\end{figure}

\begin{figure}
\includegraphics[width=135mm]{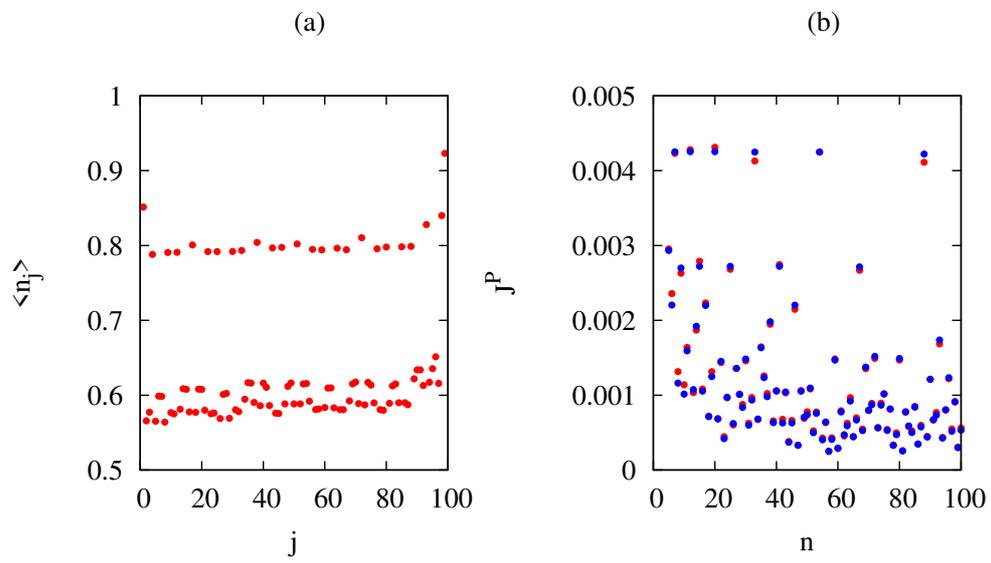}
\caption{\label{fibonacci} (Color online) Panel~(a) shows $j$
  dependence of occupation density~$n_j$ for the Fibonacci chain.
  Panel~(b) shows chain size ($n$) dependence of the particle
  current~$J^P$ for the Fibonacci chain (red: $\gamma=0.1$, blue:
  $\gamma=1$)}
\end{figure}

\begin{figure}
\includegraphics[width=135mm]{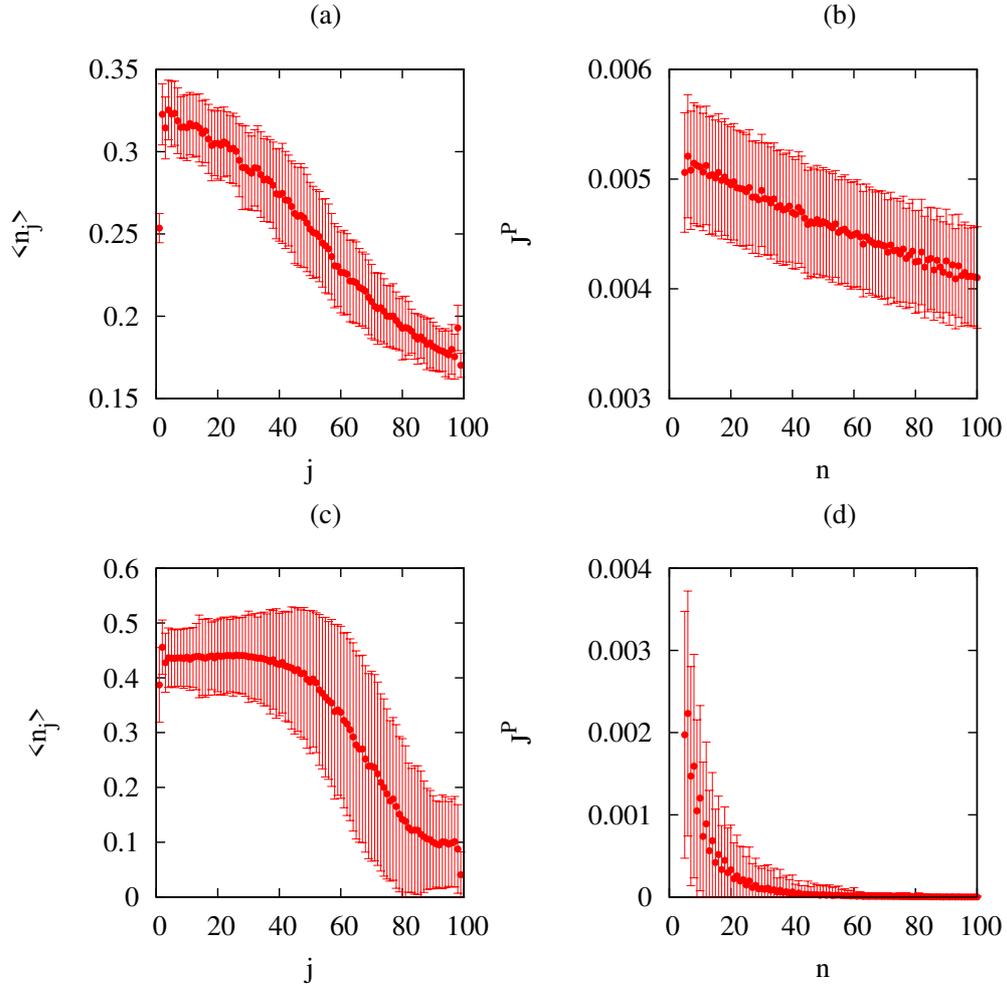}
\caption{\label{disorder} (Color online) Panel~(a) and (c) show $j$
  dependence of occupation density~$n_j$ for disordered chains.
  Panel~(b) and (d) show $n$ dependence of the particle current~$J^P$
  for disordered chains.  (a) and (c) is for $t=3,\ \delta=0.3$, and
  (b) and (d) is for $t=3,\ \delta=2.9$ }
\end{figure}

We have checked that reservoirs are close enough to equilibrium and
satisfy conservation law (\ref{particle_Landauer}).  We show the $j$
dependence of occupation density~$\ave{n_j}$ and $n$ dependence of the
particle current~$J^P$ in Fig.~\ref{periodic} (monoatomic~$t=3$ and
diatomic chain~$t_A=3,\ t_B=6$), Fig.~\ref{fibonacci} (Fibonacci
chain~$t_A=3,\ t_B=6$), and Fig.~\ref{disorder} (disordered
chain~$t=3,\ \delta=0.3, 2.9$).  For monoatomic, diatomic, and
Fibonacci chains, red dots represent $\gamma=0.1$, and blue dots
represent $\gamma=1$.

One can see that the occupation density of periodic chains is constant
except at the edge, and the particle current reaches non-zero
constant\footnote{the constant depends on the parity of system size
  $n$ for diatomic chains} for large $n$, and the transport is
ballistic.  Red dots ($\gamma=0.1$) show a small deviation since the
interaction between the system and the reservoirs is relatively small.
The Fibonacci chain shows large fluctuations in occupation and system
size dependence of the particle current, though it is very robust
against the change of $\gamma$.  For small disordered chains ($t=3,\
\delta=0.3$), the particle current decreases linearly as a function of
system size, and the occupation shows a linear profile.  For
disordered chains ($t=3,\ \delta=2.9$), occupation profile has a kink
shape and the particle current decreases nearly exponentially.

For sufficiently small
thermal and chemical gradients, the particle and the heat current,
defined as $J_Q\equiv J_E-\bar{\mu} J_P$
($\bar{\mu}=(\mu_L+\mu_R)/2$), depend linearly on the external
gradients as \cite{Domenicali54,Callen48,Groot84}
\begin{eqnarray*}
J_Q &=& L_{QQ}\Delta \beta-L_{QP} \beta \Delta \mu \ ,
\\
J_P &=& L_{PQ}\Delta \beta-L_{PP} \beta \Delta \mu \ ,
\end{eqnarray*}
where $\Delta \beta\equiv \beta_R-\beta_L$ and $\Delta \mu \equiv
\mu_R-\mu_L$.  The second law of thermodynamics imposes
definite-positiveness of the matrix of Onsager coefficients $\bf{L}$,
which implies $L_{QQ}\ge 0$ and $L_{PP}\ge 0$, and if the dynamics is
time-reversible, the Onsager's reciprocity relation $L_{PQ}=L_{QP}$
holds.

\begin{figure}
 \includegraphics[width=135mm]{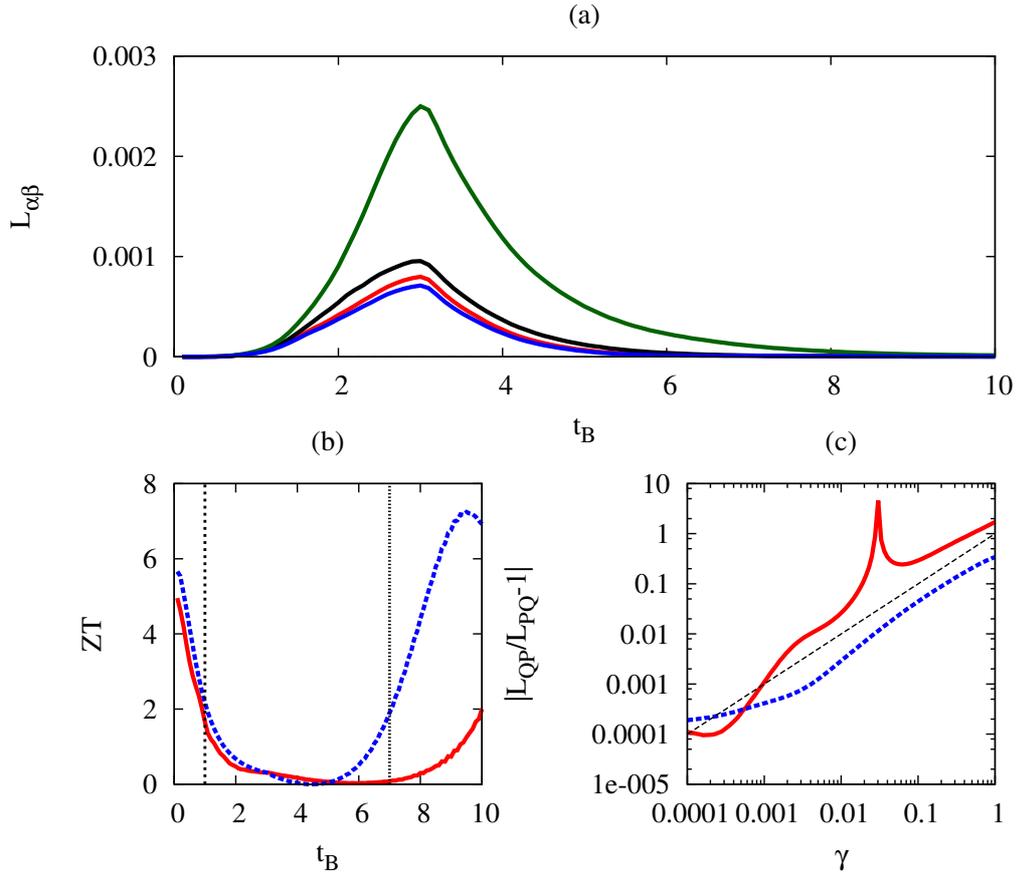}
 \caption{\label{Onsager} (Color online) Panel (a) shows $t_B$
   dependence of the Onsager coefficients (color code: $L_{QP}$ - red,
   $L_{PP}$ - black, $L_{QQ}$ - green, $L_{PQ}$ - blue) for the
   Fibonacci chains.  Panel (b) shows $t_B$ dependence of $ZT$ for
   diatomic (blue) and Fibonacci chains (red).  Hoppings $t_A$ are
   chosen as $t_A=3$.  For diatomic chains, they have exponentially
   small current outside $t_b\in [1,7]$ indicated in the figure.
   Panel (c) shows the $\gamma$ dependence of asymmetry
   $L_{QP}/L_{PQ}-1$ for diatomic (blue) and Fibonacci (red) chains
   (dashed line indicates linear growth).  $t_A$ and $t_B$ are chosen
   as $t_A=3,\ t_B=6$.}
\end{figure}

In Fig.~\ref{Onsager} we consider the Fibonacci chains with $t_A=3$
and show the dependence of various properties of $\bf{L}$ on the other
hopping parameter $t_B$.  Fig.~\ref{Onsager}(a) shows the $t_B$
dependence of all Onsager coefficients.  One sees that all
coefficients are positive, where we remark that possibility of
negative off-diagonal elements were discussed in \cite{ABMP12}.

Fig.~\ref{Onsager}(b) shows the thermoelectric figure-of-merit
$ZT\equiv L_{PQ} L_{QP}/{\rm det} \bf{L}$ \cite{Thermoelectricity} for
diatomic and Fibonacci chains.  One sees that diatomic chains have
larger ZT than the Fibonacci chains for the most parameter regimes.

Fig.~\ref{Onsager}(c) shows the $\gamma$ dependence of
$|L_{PQ}/L_{QP}-1|$ for diatomic and Fibonacci chains.  We see that
the Onsager reciprocity is roughly linearly broken by increasing
$\gamma$ for the diatomic chains.  It is very similar for the
Fibonacci chains, however here the asymmetry has a cusp shape near
$\gamma\sim 0.03$, and one should be more careful with the choice of
$\gamma$ in order to have (approximate) time-reversal symmetry.

\section{Conclusions}
We have established a conservation law for particle number, and a
conservation law for the sum of energy current along the chain and
dissipation at the boudaries of the chain.  The conservation laws are
valid for generic one-dimensional chains connected to meso-reserovirs,
which desribed by bilinear hamiltonian.

As an application, we have studied the transport properties of
monoatomic, diatomic, quasi-periodic, and disordered chains.  We have
observed wide fluctuations in the occupation and the system size
dependence of the particle current for the Fibonacci chain.  For the
disordered chain, we have observed linearly or exponentially
decreasing currents.  The occupation was shown to have either linear
profile or kink shape, respectively.

\section*{Acknowledgements}
  The authors thank J. von Delft, D. Kosov, Y. Ohta, K. Saito and
  M. \v{Z}nidari\v{c} for discussions on related subjects. SA thanks
  Fondecyt 3120254 for support.  TP acknowledges supports by the
  grants P1-0044 and J1-2208 of the Slovenian Research Agency.  TP and
  CM-M acknowledge partial support from Regione Lombardia through project
  ``THERMOPOWER''.  FB and TP thanks international collaboration
  project Fondecyt 1110144.  Finally FB and SA thanks anillo ACT 127.

\end{document}